\documentclass[pre,twocolumn]{revtex4}

\usepackage{graphicx,amssymb}

\begin{document}
\title{Human and computer learning: An experimental study}
\author{Alexandra C. Tsallis$^{a}$, Constantino Tsallis$^{b,c}$, \\
Aglae C.N. de Magalhaes$^b$ and  Francisco A. Tamarit$^d$ \footnote{xanda@rionet.com.br, tsallis@cbpf.br, aglae@cbpf.br, \\ tamarit@famaf.unc.edu.ar}}
\address{$^a$Instituto de Psicologia, Pos-Graduacao em Psicologia Social\\
Rua Sao Francisco Xavier 524, $10^o$ andar, Bloco F\\ 
Universidade do Estado do Rio de Janeiro, 20559-900 Rio de Janeiro-RJ, Brazil\\
$^b$Centro Brasileiro de Pesquisas Fisicas\\
Xavier Sigaud 150, 22290-180 Rio de Janeiro-RJ, Brazil\\
$^c$Santa Fe Institute, 1399 Hyde Park Road, Santa Fe, New Mexico 87501, USA\\
$^d$Facultad de Matematica, Fisica y Astronomia\\ 
Universidad Nacional de Cordoba, Ciudad Universitaria, Cordoba, Argentina }
\draft

\begin{abstract}
Simple memorizing tasks have been chosen such as a binary code on a $5 \times 5$ matrix.  After the establishment of an appropriate protocol, the codified matrices were individually presented to 150 university students (conveniently pre-selected) who had to memorize them. Multiple presentations were offered seeking perfect performance verified through the correct reproduction of the code. We measured the individual percentual error as a function of the number of successive presentations, and then averaged over the examined population. The {\it learning curve} thus obtained  decreases (almost monotonically) until becoming virtually zero when the number of presentations attains six. 
A computer simulation for a similar task is available which uses a two-level perceptron on which an algorithm was implemented allowing for some degree of {\it globality} or {\it nonlocality} (technically referred to as entropic {\it nonextensivity} within a current generalization of the usual, Boltzmann-Gibbs, statistical mechanics). The degree of nonextensivity is characterized by an index $q$, such that $q=1$ recovers the usual, extensive, statistical mechanics, whereas $q \ne 1$ implies some degree of nonextensivity. In other words, $q-1$ is a (very sensitive) measure of globality (gestalt perception or learning). The computer curves fit well the human result for $q \simeq 1.02$. It has been verified that even extremely small departures of $q$ from  unity lead to strong differences in the learning curve. Our main observation is that, for the very specific learning task on which we focus here, humans perform similarly to slightly nonextensive perceptrons. In addition to this experiment, some preliminary studies were done concerning the human learning of ambiguous images (based on figure-background perception).
In spite of the complexity of drawing conclusions from such a comparison, some generic trends can be established. Moreover, the enormous and well known difficulty for computationally defining semantic, hierarchic and strategic structures reveals clear-cut differences between human and machine learning. 
\end{abstract}. 

\maketitle

\date{\today}

\section{Introduction}

It is possible to consider that cognitive psychology appeared as a reaction to behaviorist approaches, where the mental content plays the role of a black box \cite{baum,miller}. In contrast, this content constitutes a central issue in cognitive sciences. Consequently, the use of computers to implement or imitate human intellectual tasks naturally emerged as a methodological tool, and even as a powerful metaphor, in the investigation of mental processes such as intelligence, learning, memorization, among others. It is along these lines that we prepared a sequence of experiments with humans, to be later on compared with similar experiments, or simulations, done with computers provided with appropriate algorithms, in particular simple perceptrons. It should of course be clear that such comparisons have philosophical implications (see, for instance, \cite{baum,penrose,button,green}), which we address in the present work only in the Conclusions (Section V).

In the present study, two different memorization tasks were implemented. The first of them was relatively simple, namely the memorization of simple binary codes in $5 \times 5$ and $ 7 \times 7$ matrices. Most of the present effort is dedicated to this analysis. The second task was, intellectually speaking, sensibly more complex. It consisted of learning ambiguous images, where the figure-background reversal  is crucial. Although semantic and strategic aspects are present in both learning tasks, the second one is by far more delicate and reveals higher level cognitive phenomena. Consistently, our study leads to quantitative information concerning the first task, whereas only some qualitative features were determined for the latter.

At the level of the comparison with computational simulations, our emphasis is put on whether learning occurs in an {\it extensive} or a {\it nonextensive} manner. These terms will be mathematically defined and further analyzed in Section III. They are currently used in statistical mechanics and thermodynamics, branches of physics dedicated to the study of the connections between the microscopic and the macroscopic worlds. At the present stage, it is enough to think of extensivity as a form of {\it nonglobality}  or {\it locality} \cite{hoffman,watters}, as opposed to nonextensivity or {\it globality} or {\it nonlocality}. In a loose manner, they respectively correspond to the {\it molecular} and {\it molar} approaches in psychology, i.e., the system is perceived as the sum of its parts, or as different from the sum of its parts. 

In Section II we present the experimental study with humans. In Section III we describe the entropic concepts that we use with regard to the computational simulations that have been implemented. In Section IV we compare both approaches, namely with humans and computers. Finally, we conclude in Section V.

\section{The experiment}

The experiment consists in individually learning fixed binary codes (represented by the signs {\bf $+$} and {\bf $\bigcirc$}) in $n \times n$ matrices. It was presented to each person, initially one $5 \times 5$ matrix (see Fig. 1) and, then, two $7 \times 7$ binary matrices (see Figs. 2 and 3). 
In order to avoid any kind of uncontrolled cultural effect, the location of the symbols $+$ and $\bigcirc$ in all the matrices was randomly generated by computer and fixed once for ever.           
It was carried out with approximatively 150 university students whose specialty was not directly related to geometry, mathematics or images, in order to avoid professional bias. For instance, students of physics, mathematics, engineering, architecture, visual publicity were excluded. The experiment population was mainly constituted by students of psychology, administration, social service and social comunication of the Federal University of Rio de Janeiro. Approximately 30 students were used for preliminary tests in order to fix an optimal experimental protocol (matrix sizes, binary code of each matrix, exhibition and hiddening times, among others). Then, precisely 120 (92 female and 28 male) students were exposed to the same protocol, and only their results were taken into account for quantitative purposes in the statistical processing (averaging, in particular). Their ages ranged between 17 and 27 years old, the majority of them being around 22 years old. 

Each of them was sequentially isolated in a peaceful room and tested, for about 40 minutes, by one of us.
Information about the scope of the research was given to each one of the 120 individuals . It was told that the experiment was not measuring their intelligence (so that they would feel relaxed), and that it was important for understanding how learning occurs (so that they would seriously try to perform satisfactorily). After these instructions were given to each one of the 120 students, three matrices were shown. In all cases, a $5 \times 5$ matrix was first shown, one of the two complementary matrices (i.e, obtained by interchanging the symbols {\bf +} and {\bf o}). Then, one of the four $7 \times 7$ matrices was shown (matrix in Fig. 2 to 30 students, matrix in Fig. 3 to other 30, and so on). Then the other noncomplementary $7 \times 7$ matrix was shown. The sequence of each individual test was as follows:\\
{\it (1)} after some general explanations, an empty $5 \times 5$ matrix and the $\bigcirc$ and $+$ symbols were shown, and the subjects were asked to randomly fill the matrix with the two symbols. This step aimed to provide to the individuals some familiarity with the experiment;\\
{\it (2)} then Fig. 1 was exhibited during 8 seconds and then hidden. The student had to try to reproduce it on the spot on an empty matrix. When this was done, the operation was repeated after a rest interval of 10 seconds. The $5 \times 5$ matrix was never shown more than 10 times (the individuals started feeling tired after 10 times). The matrix was considered to be learnt if no error was made after two successive exhibitions;\\
{\it (3)} then the learning test was repeated by successively using two of the four $7 \times 7$ matrix, one at a time;\\
{\it (4)} finally, the student was asked  to briefly describe how he(she) proceeded to learn. 

The {\it error} $\epsilon_i(t)$ of the $i$-th individual ($i=1,2,3,...,120$; $t=0$ corresponds to the initial random filling) is defined as the number of wrong elements of the filled $n \times n$ matrix ($n=5,7$); $0 \le  \epsilon_i(t) \le n^2$. Typical results are presented in Figs. 4, 5 and 6 for the $5 \times 5$ and in Figs. 7 and 8 for one of the four $7 \times 7$ matrices (the figures associated with the other three $7 \times 7$ matrices are quite similar in fact). Although in a quite different context, results that have some connection with the present ones have been exhibited in \cite{taatgen}. 

The averages $\langle \epsilon \rangle(t) \equiv \frac{1}{N}\sum_{i=1}^N \epsilon_i(t)$ (with $N=120$ if the entire set of students is used for averaging) are shown in Fig. 5. These curves already achieved the aspect presented in this figure when the averages were performed with approximately 80 individuals. Using the entire set of 120 results just improved the precision but incorporated no new qualitative elements in the curves. If we define the {\it learning time} as the number of times shown before learning, excluding those who did not suceed learning that particular matrix untill the end of the experiment, we can check that it is of the order of 6. 

Incidentally we verified an interesting (and indeed unexpected) cultural phenomenon. For the $5 \times 5$ matrix we were expecting $\langle \epsilon \rangle(0)$  to be close to $25/2=12.5$ since, before starting to show the codified matrix, we asked to {\it randomly} fill the empty matrix with symbols {\bf o} and {\bf +}, indicated in {\it this} order above the empty matrix, if we look at them from left to right. In variance with this reasonable expectation, we found, for the first 60 students, $\langle \epsilon \rangle(0) \simeq 14$. 
After some hesitation about what could be the cause of this asymmetry (e.g., could it be the different semantics humans associate with a circle or a cross?), we speculated that it could be the fact that Portuguese language (within which the Brazilian population is educated) is read {\it from left to right}. Then, to the second and last set of 60 students, an empty Fig. 1 was presented with the symbols in the ordering {\bf + o} instead of {\bf o +}. Very symptomatically, we then obtained $\langle \epsilon \rangle(0) \simeq 11.5$, the overall average  being consequently $(14 + 11.5)/ 2 = 12.75$, reasonably close to 12.5  as initially expected! To confirm this cultural cause of the observed asymmetry, it would be interesting to repeat the experiment with say arabic students (educated within {\it right-to-left} reading).

Another interesting feature that we observed is that, for many individuals, $\langle \epsilon \rangle(2) >  \langle \epsilon \rangle(1)$, thus systematically contradicting the overall monotonic tendency of $\langle \epsilon \rangle(t)$ to decrease with time. The reason for this kind of a priori unexpected behavior appeared to be (as commented by the individuals themselves during the free final conversation) that, after seeing for the first time the code to be learnt, the student dedicated a good part of his (her) attention to establish a``strategy" for learning rather to properly learn the matrix. Let us mention, by the way, that in an experiment like the present one it is quite hard to differentiate between learning  the strategy and memorizing the matrix within that strategy. This kind of modelization is supported by the fact that, several months after conclusion of the experiment, quite a few students still remembered the strategy, while they had completely forgotten the particular matrix code itself. 

Let us now briefly comment the second experiment we developed.
We chose several ambiguous images, all of them being susceptible of two mutually excluding interpretations on the basis of figure-background reversal. For example, if one sees in Fig. 15 \cite{ambiguous} a young woman, one does not simultaneously see the old woman, and reciprocally . We implemented this more complex learning task by showing the image and then asking to the individual what he sees. Then we tried to count the time needed by the person in order to recognize the other image interpretation. This perception mechanism is sometimes referred to as {\it reversal} of the figure-background. It turned out that the times involved in this type of experiment, and very specifically the slowness of learning, if any, how to recognize the reversal, were so ill-defined that we decided not to proceed with this protocol. The question of how to conveniently quantify such learning remains, therefore, an open question (see also \cite{maturana,arendt}).

\section{Entropic concepts and computational simulations}

A great variety of computer learning algorithms are available in the literature. It is clear that all of them process, in one way or another, information. Entropy is well known to be a convenient tool for quantifying (lack of) information. It can therefore be used in the context of any learning algorithm, at least in principle. We shall use it here in connection with the specific perceptron we shall describe later on. For convenience, let us briefly review at this point some basic notions  
about the entropic forms we are referring to in the present paper. The Boltzmann-Gibbs-Shannon (BGS) entropy is the basis of standard statistical mechanics and thermodynamics. It is defined (in its discrete version) by \cite{huang}
\begin{equation}
S_{BGS} = - \sum_{i=1}^W p_i \ln p_i\;,
\end{equation}
where $W$ is the number of microscopic possibilities accessible to the system, and $\{p_i\}$ are the associated probabilities ($\sum_{i=1}^W p_i =1$); for simplicity, we have taken Boltzmann constant equal to unity. This entropy becomes maximal at equiprobability, i.e., $p_i=1/W$ for all $i$, and achieves the value 
\begin{equation}
S_{BGS} = \ln W\;,
\end{equation}
which is the celebrated Boltzmann formula. 

If we consider a composite system made of two (probabilistically) independent systems $A$ and $B$, i.e., if we assume that $p_{ij}^{A+B} = p_i^Ap_j^B$, and replace this into Eq. (1), we straightforwardly obtain
\begin{equation}
S_{BGS} (A+B) = S_{BGS}(A) + S_{BGS}(B)\;,
\end{equation}
which can be phrased as {\it the entropy of the whole is the sum of the entropies of the parts}. The entropic form (1) is the basis of standard, Boltzmann-Gibbs (BG), statistical mechanics and thermodynamics, and property (3) is known as {\it extensivity} or {\it additivity}. This entropy, as well as others, are in some sense  ubiquitous. Indeed, they emerge in a great variety of discussions. For example, they 
have often been  used concerning complex phenomena such as the organization of living matter (see, for instance, \cite{atlan}),
as well as other types of organization, including that of knowledge (e.g., memorization and learning), economics, linguistics, to mention but a few (see, for instance, \cite{tsallis,review}).  Before addressing the generalization of $S_{BGS}$ we are interested in here, let us mention that optimization of Eq. (1) in the presence of a constraint of the type $\sum_{i=1}^W p_i E_i = constant$ (where $E_i$ might be say microscopic energy levels), leads to
\begin{equation}
p_i \propto e^{-\beta E_i} \;,
\end{equation}
where $\beta$ is a parameter to be determined through the value of the constraint. In statistical mechanics, Eq. (4) is in fact the celebrated Boltzmann-Gibbs weight. 

In 1988, one of us (CT) proposed \cite{tsallis} the generalization of BG statistical mechanics on the basis of a more general entropic form, namely
\begin{equation}
S_q = \frac{1- \sum_{i=1}^W p_i^q}{q-1} \;,
\end{equation}
$q$ being any real number. We can verify that $S_1=S_{BGS}$, in other words, the BG formalism becomes now the $q=1$ particular case of this more general formalism. If we assume, once again, two independent systems $A$ and $B$, we can prove that
\begin{equation}
S_q (A+B) = S_q(A) + S_q(B)+(1-q)S_q(A)S_q(B)\;,
\end{equation}
which can be phrased as {\it the generalized entropy of the whole is different from the sum of the generalized entropies of the parts}. This property is referred to as {\it nonextensivity} or {\it nonadditivity}. To be more precise, if we take into account that $S_q$ is always zero or positive, Eq. (6) implies that $S_q (A+B) > S_q(A) + S_q(B)$ if $q<1$ and $S_q (A+B) < S_q(A) + S_q(B)$ if $q>1$. Only if $q=1$ we have that $S_q (A+B) = S_q(A) + S_q(B)$. It is from property (6) that the terms {\it nonextensive} statistical mechanics and thermodynamics have been coined (for reviews see \cite{review}). Analogously to what we did before, if we optimize $S_q$ in the presence of the constraint $\sum_{i=1}^W p_i^q E_i / \sum_{i=1}^W p_i^q = constant$, we obtain \cite{tsallis}
\begin{equation}
p_i \propto [1-(1-q)\beta^\prime E_i]^{1/(1-q)} \;,
\end{equation}
where, as before,  $\beta^\prime$ is a parameter to be determined through the value of the constraint. Notice that for the normal regime for $\beta^\prime$, i.e., $\beta^\prime>0$, we have, for large values of $E_i$, a long {\it power-law} tail for $q>1$, whereas we have a short {\it exponential} tail for $q=1$;  finally, for $q<1$, we have a cutoff. 
Eq. (7) can be re-written in the Boltzmann-Gibbs form, namely
\begin{equation}
p_i \propto e^{-\beta^\prime E_i^\prime} \;,
\end{equation}
where
\begin{equation}
E_i^\prime \equiv \frac{-1}{\beta^\prime (1-q)} \ln [1-(1-q) \beta^\prime E_i] \;.
\end{equation} 
In other words, in what concerns the optimizing distribution, $E_i^\prime$ plays the role of an effective energy which replaces $E_i$ (in the $q \to 1$ limit, of course we recover $E_i^\prime \to E_i$). This effective energy can be used to $q-$generalize a variety of microscopic and mesoscopic equations. One such example is the Langevin equation, on which the perceptron that we use here has been constructed.

\begin{figure}
\begin{center}
\includegraphics[width=6cm,angle=0]{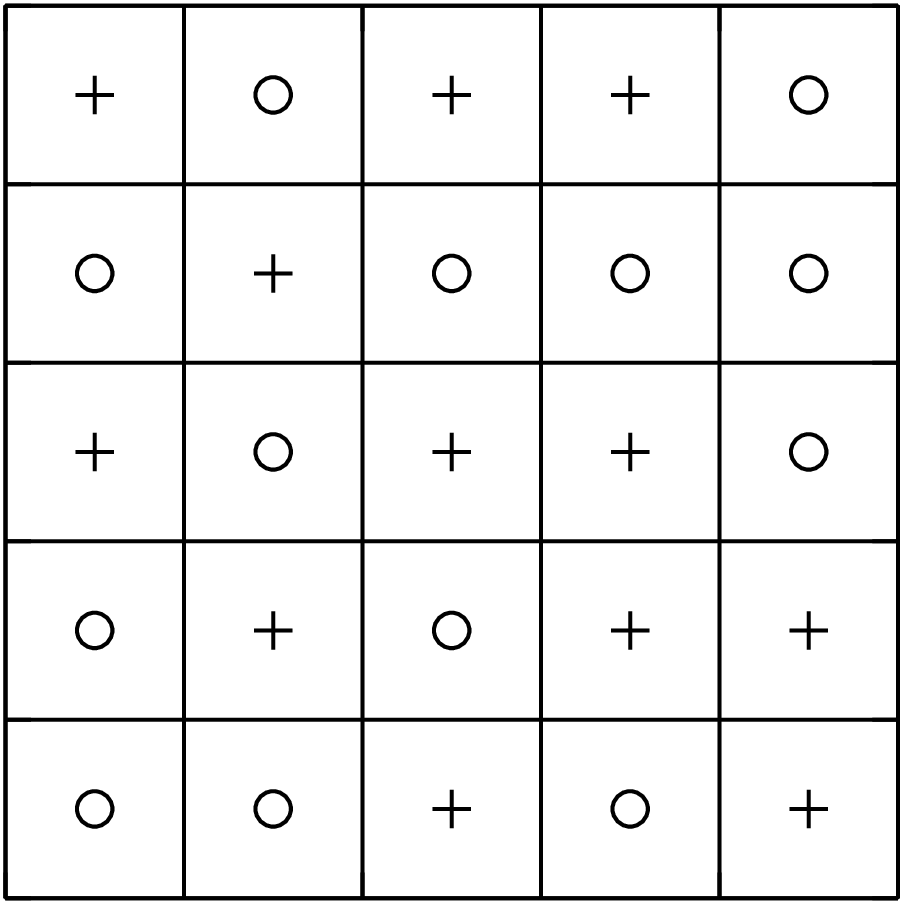}
\end{center}
\caption{\small The $5 \times 5$ matrix.} 
\end{figure}

\begin{figure}
\begin{center}
\includegraphics[width=5.5cm,angle=0]{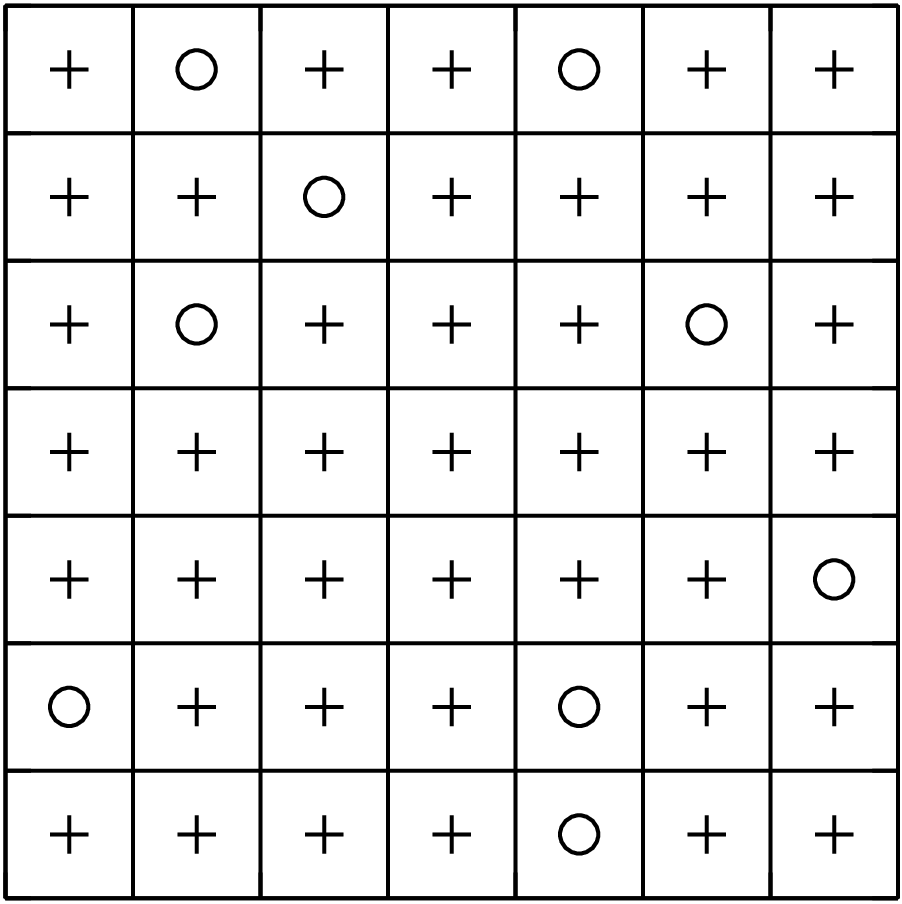}
\end{center}
\caption{\small First $7 \times 7$ matrix. This matrix is referred to as the $+$diagonal matrix. Its dual matrix is obtained by permutating the $+$ and the $\bigcirc$ symbols, and is referred to as the $\bigcirc$diagonal matrix.} 
\end{figure}

\begin{figure}
\begin{center}
\includegraphics[width=5.5cm,angle=0]{figure3.eps}
\end{center}
\caption{\small Second $7 \times 7$ matrix.   This matrix is referred to as the $\bigcirc$column matrix. Its dual matrix is obtained by permutating the $+$ and the $\bigcirc$ symbols, and is referred to as the $+$column matrix.   } 
\end{figure}

\begin{figure}
\begin{center}
\includegraphics[width=9cm,angle=0]{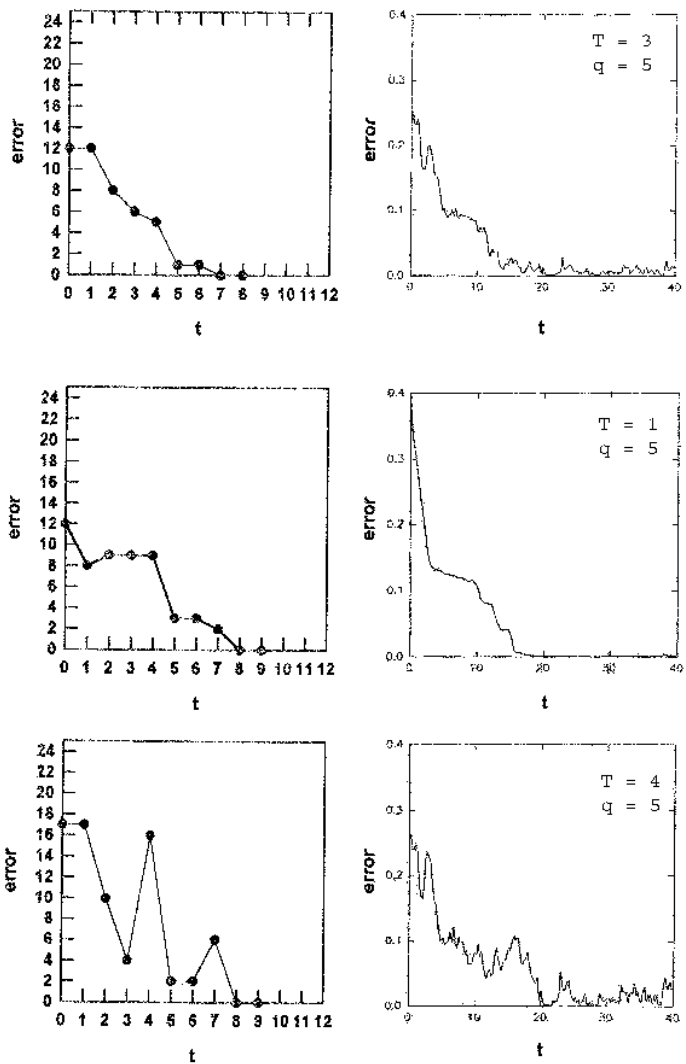}
\end{center}
\caption{\small Typical examples of time evolution of the error associated with the $5 \times 5$ matrix. The three left correspond to three different individuals (the abscissa is the number of presentations of the matrix; the ordinate is the number of matrix elements incorrectly reproduced). The three right (see [16] for further details) correspond to three different initial conditions for the perceptron (the abscissa is the number of iterations; the ordinate is the percentual error).} 
\end{figure}

\begin{figure}
\begin{center}
\includegraphics[width=8cm,angle=0]{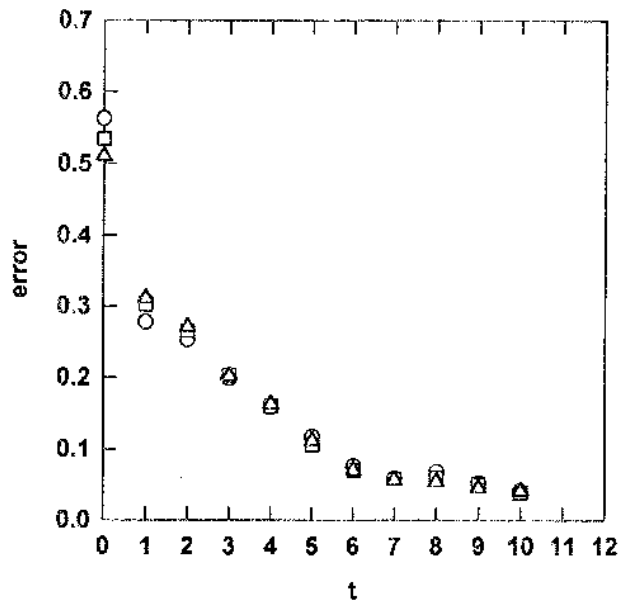}
\end{center}
\caption{\small Average error associated with the learning of the $5 \times 5$ matrix as a function of the number of presentations: circles, squares and triangles respectively correspond to averaging over $73$, $92$ and $120$ individuals.} 
\end{figure}

\begin{figure}
\begin{center}
\includegraphics[width=7cm,angle=0]{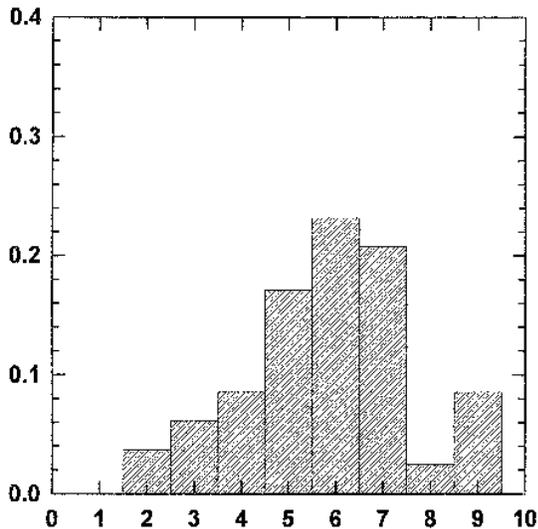}
\end{center}
\caption{\small Histogram of the probabilities corresponding to the $5 \times 5$ matrix. The abscissa is the ordinal of the presentation at which the individual learnt the matrix: 82 (out of 120) individuals learnt the matrix before or at the 9th presentation (we recall that the 10th presentation was used to confirm the learning at the 9th one); 38 individuals did not succeed (and are not computed in the histogram).} 
\end{figure}

\begin{figure}
\begin{center}
\includegraphics[width=7cm,angle=0]{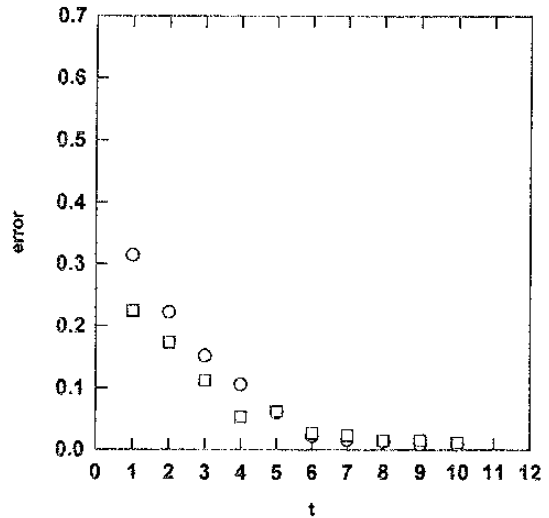}
\end{center}
\caption{\small Average error associated with the learning of the  $7 \times 7$ $+$diagonal matrix as a function of the number of presentations: the circles correspond to averaging the results of $30$ individuals to whom the $+$diagonal matrix was shown in {\it first} place;  the squares correspond to averaging the results of $30$ individuals to whom the $+$diagonal matrix was shown in {\it second} place (after having seen, in {\it first} place, the $\bigcirc$column matrix). } 
\end{figure}

\begin{figure}
\begin{center}
\includegraphics[width=7cm,angle=0]{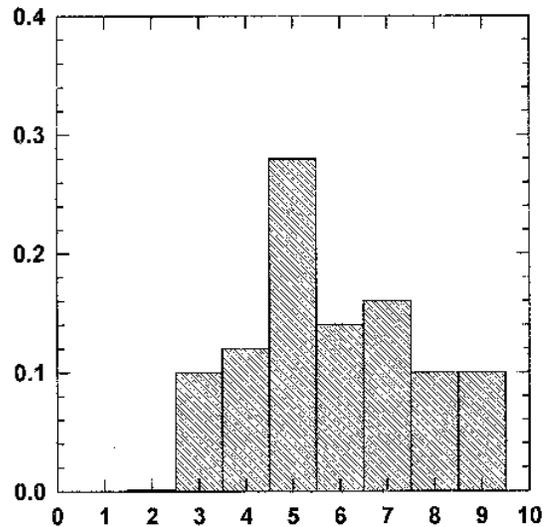}
\end{center}
\caption{\small Histogram of the probabilities corresponding to the {\it first} shown $7 \times 7$ matrix, either the  $+$diagonal or the $\bigcirc$diagonal ones.  The abscissa is the ordinal of the presentation at which the individual learnt the matrix: 50 (out of 60) individuals learnt the matrix before or at the 9th presentation (we recall that the 10th presentation was used to confirm the learning at the 9th one); 10 individuals did not succeed (and are not computed in the histogram).} 
\end{figure}

\begin{figure}
\begin{center}
\includegraphics[width=6cm,angle=0]{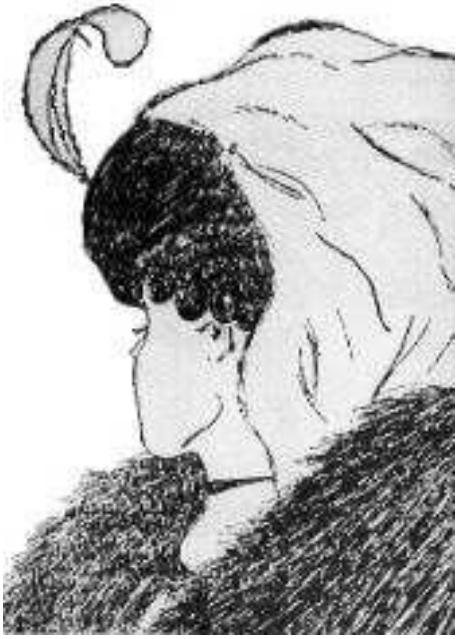}
\end{center}
\caption{\small The well known ambiguous figure ``My Wife and My Mother-in-law", created by W.E. Hill in 1915, and originally published in {\it Puck}.}

\end{figure}

\begin{figure}
\begin{center}
\includegraphics[width=8.5cm,angle=0]{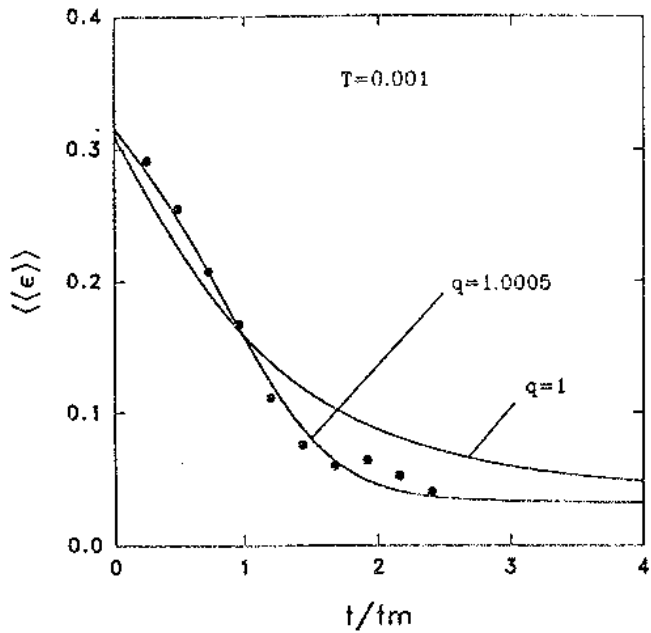}
\end{center}
\caption{\small Typical average error curves for the perceptron ($5 \times 5$ matrix) as a function of the number of iterations ($T$ is a parameter of the perceptron). The perceptron has been chosen with $5 \times 5 = 25$ binary inputs, in order to simulate the human task (dots) as closely as possible on the average (taken on 92 individuals in this example). See [16] for further details.} 
\end{figure}

\begin{figure}
\begin{center}
\includegraphics[width=8.5cm,angle=0]{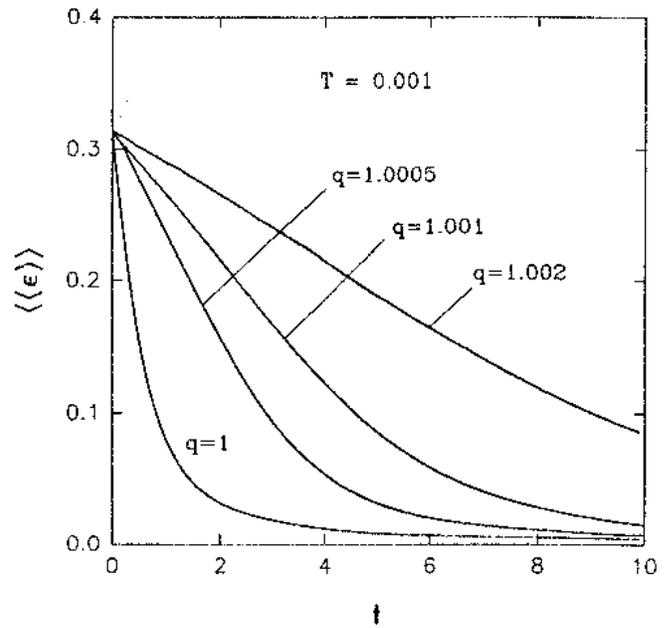}
\end{center}
\caption{\small Typical average error curves for the perceptron ($5 \times 5$ matrix) as a function of the number of iterations ($T$ is a parameter of the perceptron).  We notice the extreme sensitivity to the value of $q$ in the neighborhood of $q=1$. See [16] for further details.} 
\end{figure}

\begin{figure}
\begin{center}
\includegraphics[width=8.5cm,angle=0]{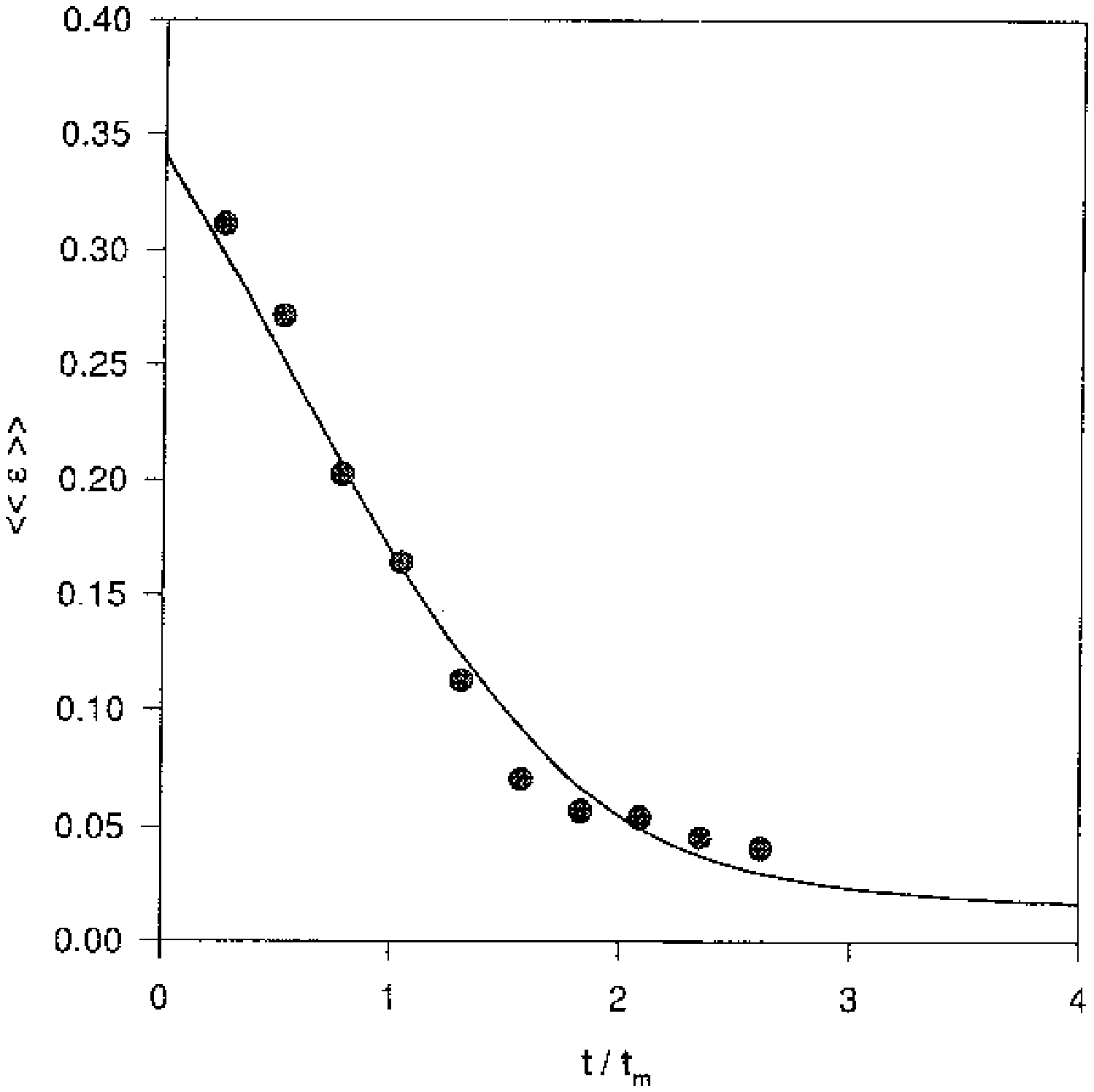}
\end{center}
\caption{\small Time evolution of the error for the $5 \times 5$ matrix, where $t_m$ is the value of $t$ at which the error becomes half of its value at $t=0$. The dots correspond to averaging the experimental data  with $120$  humans. The continuous curve has been obtained averaging a large number of initial conditions for the perceptron with the gain parameter $g=0.91$, the temperature-like parameter $T=0.02$ and the entropic index $q=1.02$. It is with these values that optimal fitting was obtained for the experimental data. The parabolic extrapolation of the experimental data corresponding to $t=1, \,2$ and $3$ provides, for $t=0$, the value of $0.316$ for the percentual error, which corresponds to $0.316 \times 25= 7.9$ for the absolute error. Therefore the value of $t_m$ corresponds to the value of $t$ at which the absolute error equals $7.9/2=3.95$. Since no integer value of $t$ corresponds exactly to this value, a linear interpolation has been performed, yielding $t_m=4.2$ for the experimental data with 120 individuals. See [16] for further details.} 
\end{figure}

As we see, the entropic property (6) has a kind of {\it gestalt}-like flavor. Its use constitutes a natural choice if we desire to deal with informational phenomena involving global or nonlocal aspects. Since this might well be the case of human learning, we have adopted this formalism in order to have the possibility of comparing human and machine learnings.

To do so, a nonextensive perceptron has been implemented \cite{cannasetal} which performs a task similar to the learning of the $5\times5$ and $7\times7$ matrices that were exposed to the students, according to the experimental protocol we described earlier. In order to perform calculations, the perceptron needs an internal dynamical equation. To fulfill this requirement, the $q$-generalized Langevin equation previously introduced by Stariolo \cite{stariolo} was implemented in the perceptron.  Some typical runs of the perceptron are shown in Fig. 4, and typical averages are shown in Figs. 16, 17 and 18 (from \cite{cannasetal}). 

\section{Comparison of the human and computer results}

The purpose of the present section is to compare the results obtained with humans and those obtained with the nonextensive perceptron. The comparison will be illustrated on the learning/memorizing of the $5 \times 5$ matrix. We shall verify that, for this specific task, the human and perceptron results can be amazingly similar. This can be checked on Fig. 4. The three individual results (on the left) are indeed similar to the perceptron realizations with three different initial conditions (on the right). The three human examples have been chosen as to exhibit typical cases. The three perceptron examples have been chosen in order to have an overall aspect similar to the human ones.

Averages over many realizations of the  data just presented  are shown (with dots) in Figs. 16 and 18. We have rescaled the time variable in such a way that comparison becomes possible on the same graph. More precisely, we have expressed time in units of the corresponding half-time $t_m$, defined as the value of time at which the average error curve decays to its half value. A rescaling such as this one is clearly necessary in order to quantitatively compare the results. Indeed, human ``time" is here represented as the number of presentations, whereas perceptron ``time" essentially corresponds to the number of computer iterations. These two numbers being of a completely different nature (see also \cite{hoffman}), it is clear that rescaling becomes necessary.

We may consider Fig. 18 as the central result of the present work. We verify that, for the specific task of learning/memorizing 25 binary states (on a matrix for the humans), humans and machines are remarkably similar. Of course, the parameters of the perceptron have been chosen in such a way as to optimize the overall fitting to the human data. The number of individuals that have been averaged is 120, and we have verified that no sensible variation is obtained under increase of that number. In the language of statistical mechanics, we may say that we have practically attained the thermodynamic limit. There is a little bump in the human results at $t/t_m \simeq 2$: we have not identified its origin (perhaps fatigue of the tested individuals, perhaps something else) . The perceptron does not exhibit such bump. As we see, the perceptron that fits the data has some degree of nonextensivity, as was conjectured in the beginning of the present work. Although $q$ is quite close to unity, we must take into consideration the fact that the error curves have been shown to be (see Fig. 17) extremely sensitive to the degree of nonextensivity in the neighborhood of $q=1$ (extensive case).

\section{Conclusions}

In summary, we have implemented an experimental study aiming to measure the learning/memorizing performance of humans on simple codes on matrices. It is on purpose that we simultaneously use the words ``learning" and ``memorization". Indeed, the experiments clearly showed that the improvement of correct answers was due to a mixture of memorization of the specific codified example and devising learning {\it strategies} (symmetry rules and other mnemonic tricks) in order to efficiently implement the memorization effort. In fact, after several months, we informally verified that the individuals had forgotten the codes, but still remembered the strategy they used for memorizing them. 

Through comparison with machine results, we verified that this particular human task was executed with clear indication of a slight, though very efficient nonextensivity (or globality) quantified by the entropic index $q$. This index appeared to be slightly {\it above} unity, which characterizes {\it slower} learning/memorizing (the error curves takes longer to become basically zero), but perhaps higher ability for devising strategies. It is then allowed to conjecture that human nature evolved, during successive generations, not so much to strongly improve the speed associated with such kind of memorization, but rather to improve the capacity of spontaneously and quickly generating intellectual strategies for performing tasks such as memorization.   

We also applied a similar experimental protocol for learning/memorizing how to analyze complex figure-background images in order to quickly realize alternative (typically two)  interpretations of the figures. We verified  that, unless much more sophisticated experimental protocols and computational algorithms are deviced, cognitive tasks with important semantic content are by all means nontrivial to measure and compare. For such complex tasks, even more than for the simple binary learning/memorization addressed here, the role of strategies might well be fundamental, although this remains to be proved. On more general grounds, the scenario which emerges is that nonextensivity seems to serve to humans for achieving {\it abduction}, one of Charles Sanders Pierce three basic forms of inference (see \cite{pierce} and references therein). In other words, given its intrinsic {\it nonlocal} nature (strong collective correlations are necessary for making the entropic index $q$ to differ from unity), it is plausible that nonextensivity constitutes the structure necessary to make {\it metaphors}. Given the very high intellectual level attributed, since Aristotle, to metaphors \cite{aristotle}, it is allowed to think that it has some specific relation with the nature of the one that we might consider as the {\it animal who makes metaphors}. We may use {\it ``homo metaphoricus"} to express this concept. 
Further developments, on both philosophical and cognitive-psychological grounds, within the frame that we have outlined here would naturally be very welcome. They could reinforce or exclude the interpretation that our present  human-machine comparison suggests. 

\section*{Acknowledgements}

We thank A.B. Lima and K.B. Miziara for assistance during the early stages of the present work, as well as S.A. Cannas and D.A. Stariolo for making available to us their perceptron curves. Computational assistance and useful remarks by L. Silva, C. Anteneodo, F. Baldovin and M.P. Albuquerque are acknowledged as well. We finally thank CAPES, CNPq, PRONEX and FAPERJ (Brazilian agencies) for partial financial support.

\end{document}